\documentclass[aps,twocolumn,showpacs,preprintnumbers,amsmath,amssymb]{revtex4}

\usepackage{graphicx}
\usepackage{dcolumn}
\usepackage{bm}
\usepackage{amsmath}

\begin{document}

\preprint{APS/123-QED}

\title{INFLUENCE OF RANDOM BULK INHOMOGENEITIES ON QUASI-OPTICAL CAVITY RESONATOR SPECTRUM}

\author{E.\,M. Ganapolskii}
\author{Z.\,E. Eremenko}
 \email{zoya@ic.kharkov.ua}
\author{Yu.\,V. Tarasov}
\affiliation{%
Institute for Radiophysics and Electronics, National Academy of Sciences of Ukraine,\\
12 Proscura St., 61085 Kharkov, Ukraine }


\begin{abstract}
We suggest the statistical spectral theory of oscillations in
quasi-optical cavity resonator filled with random inhomogeneities.
It is shown that inhomogeneities in the resonator result in
intermode scattering leading to the shift and broadening of spectral
lines. The shift and broadening of each line essentially depends on
frequency distance to adjacent spectral lines. With increasing the
distance the influence of inhomogeneities sharply reduces. The
solitary spectral lines which have the distance to the nearest lines
quite large is slightly changed due to small inhomogeneities. Owing
to such selective influence of inhomogeneities on the spectral lines
the effective spectrum rarefaction appears. Both the shift and
broadening of spectral lines as well as spectrum rarefaction in
quasi-optical cavity millimeter wave resonator were detected
experimentally. We found out that inhomogeneities result in
stochastization of the resonator spectrum in that mixed state
appears, i.e.  the spectrum acquires both regular and random parts.
The active self-oscillator system based on the inhomogeneous
quasi-optical cavity  millimeter wave resonator with Gunn diode was
studied as well. The inhomogeneous quasi-optical cavity millimeter
wave resonator (passive and active) can serve as a model of
semiconductor quantum billiard. Based on our results we suggest
using such billiards with spectrum rarefied by random
inhomogeneities as an active system of semiconductor laser.
\end{abstract}



\maketitle

\section{\label{sec:level1} Introduction  }
The problem of electromagnetic wave propagation in random
inhomogeneous media has been the issue of the day for several
decades. Numerous publications are devoted to the analysis of
different aspects of this problem (see [1,2] and references
therein). In this scientific area the subject of research is
normally the scattering of electromagnetic (acoustic) waves in
unbounded or partially confined  systems (for example, in
waveguides) which contain random inhomogeneities. Researches on wave
propagation in open statistically unregulated systems are stimulated
by numerous applications for long-distance signal transmission in
both radio and optical wave ranges.

The problem of electromagnetic oscillations in confined systems with
random inhomogeneities has significant meaning as well. At the same
time, by now it has not been developed enough both from theoretical
and experimental side. It is caused by, in particular, modern
theoretical studies of electromagnetic wave propagation and
scattering in random inhomogeneous media use assumption about the
isotropy and homogeneity scattering medium filled confined systems
[1,2]. Therefore, they cannot be applied in general to the study of
confined systems such as cavity resonators. The experimental study
also requires new approach to study electromagnetic oscillations in
inhomogeneous cavity resonators in wide wavelength band.

In our recent paper [3] we suggested a new spectral approach to
study confined systems with random inhomogeneous. There spectral
properties of cavity spherical quasi-optical millimeter wave
resonator filled with random sapphire particles were considered. The
sapphire particles with dimensions of the order of operating
wavelength affect significantly the resonator spectrum. The
spherical frequency degeneration is completely removed and the
spectral lines thus have chaotic distribution on the frequency
scale. The lines became wider and the quality factor is
correspondingly decreased.  The analogous broadening of spectral
lines caused by random inhomogeneities has been detected in Refs.
[4,5], where resonators with random rough boundaries were studied.

Recently the question about the influence of random inhomogeneities
on the resonator spectrum has attracted a great attention in
relation to the design of lasers on open micro-resonators [6]. In
such resonators, the whispering-gallery modes with super-high
quality factors can be excited. The possibility to realize such
quality factors depends essentially on the number of inhomogeneities
(roughness level) on the resonator boundaries.

It should be realized that cavity resonator with extremely small
dissipation loss is the almost Hermitian system whose oscillations
are formed due to the restricted motion of electromagnetic waves.
Therefore, the resonator spectrum is discrete. Random
inhomogeneities whether bulk or on the resonator boundaries not
introducing additional dissipative loss in general have not to
broaden the spectral lines. Nevertheless, the broadening of lines in
the spectrum effects and its stochastization have been detected in
experiment [3].

In this connection the question arises: what is the physical
mechanism of spectral lines broadening and spectrum stochastization
in the random inhomogeneous resonator? The study of this question is
one of the reasons of the present paper. Another motivation is
relevant to the nanoelectonics problem. Recently the great emphasis
is the study of a new type of nanoelectron systems (so-called
zero-dimensional ones). The charge carrier motion in these systems
has space restriction in all three dimensions. The peculiar example
of zero-dimensional system is a quantum dot (QD). QD is a
semiconductor area by the size of order of 10 nm with electron
(hole) conductivity and restricted from outer area by potential
barrier. Owing to finite charge carrier motion the energy spectrum
in QD is discrete and the number of spectral levels due to small QD
size is relatively few.

Recently an elegant method of QD array implementation has been
designed. This method is based on the self-organization effect in
strained double GaAs heterostructures [7-9]. The usage of QD array
as an active medium permits to design lasers with high performance
[10-12]. However, the self-organization process of QD array forming
is difficult to manage. The random inhomogeneities that usually
exist in heterostructure essentially affect on this process.  They
lead to inhomogeneous broadening of spectral lines [13] and,
correspondingly, to degradation of laser radiation quality.

The QD design method based on the self-organization effect is, in
fact, an alternative to electron lithography whose level of
development does not permit to implement the ordered array of
approximately the same QDs with small enough dispersion of their
sizes. In the present paper we propose another way to design
semiconductor laser system. This way presupposes the usage of the
same regular microscopic areas ordered array that can be implemented
by lithography as an active laser medium.  This proposal is based on
the following. At present technology of GaAs mono-crystal with super
high mobility and big length of phase coherence of charge carriers
that achieves 10 $\mu m$ and more is well-developed. Due to that the
system of microscopic regular areas with potential barrier on the
margin of each of them made from such materials can be implemented.
Because of big length of phase coherence, carriers will take part in
ballistic motion and reflect back from area margins. The motion of
charge carriers is similar to dynamics of billiard systems. Since
this motion is described by Shrodinger equation such a billiard
system can be characterized as a quantum billiard (QB).

Due to finite motion of charge carrier, electron spectrum of QB is
discrete. At the same time it is quite dense, because of a QB size
is much bigger than the wavelength of quasi-particles. This fact
makes complicate the QB usage as an active system for the
semiconductor laser, because dense spectrum decreases frequency
stability of laser radiation. The frequency jumps appear easily at
the small deviation of control parameters in the laser resonator
with dense frequency spectrum. As a result there is a problem of
"rarefaction" of the spectrum: \emph{can the dense QB spectrum be
done much sparser without changing of QB geometrical parameters?}

A quasi-optical cavity resonator has dense and discrete frequency
spectrum as well. The Maxwell equation describing electromagnetic
oscillations in such a resonator coincides with corresponding scalar
Shrodinger equation at definite conditions. All that gives an
opportunity to use the quasi-optical resonators as model objects to
study spectral properties of QB. Similar modeling using microwave
resonators was done earlier for study of the phenomena that is
relevant to quantum chaos [14-17].

In the present paper the nature of spectral lines broadening and
their shifting are studied in a quasi-optical millimeter wave cavity
resonator with random inhomogeneous inside. To clarify the mechanism
of broadening and shifting the original statistical spectral theory
based on the separation mode technique was worked out. This
technique was designed earlier by one of the author for open
waveguide systems [18,19]. Using this technique we found out that
the mechanism of spectral lines broadening was caused by intermode
scattering. At that the largest broadening are subjected to closely
set lines. And solitary lines keep high quality factor and intensity
in the random inhomogeneous resonator. It results in so-called
spectral "rarefaction".

The systematic experimental spectral measurements of numerous
realizations of quasi-optical random inhomogeneity cavity resonator
filling were carried out for theory results implementation. The
experimental results proved that the main mechanism of broadening
and shifting of spectral line is inter-mode scattering caused by
inhomogeneities resonator filling. We detected theoretically
predicted effect of "rarefaction" of the resonator spectrum. Thus,
we determined that the influence of inhomogeneities on the resonator
spectrum has both, as known, negative nature (broaden lines) and a
positive one. This positive influence is essential "rarefaction" of
the spectrum. This fact can be important in QB usage as an active
system in a semiconductor laser approach. Here we state that the
high quality modes are not subjected, practically, to broadening and
can satisfy self-excitation laser condition even under big
inhomogeneities in an active resonator medium. Exactly the stable
laser generation can be implemented under one of such modes.

Another aspect of influence of inhomogeneities on resonator spectrum
relates to its stochastization. In the paper we carried out
statistical analysis of the resonator spectrum with a different
number of inhomogeneities. We found out that the inter-frequency
(IF) interval distribution in the resonator spectrum is close to
Poisson distribution in a small number of inhomogeneities. This kind
of distribution has systems with non-correlated IF intervals. The
stochastic spectrum part appears at the increase of the number of
inhomogeneities. The spectrum is mixed, i.e. contains regular and
stochastic parts.  We estimated relationship between regular and
stochastic spectrum parts by statistical distribution analysis of IF
intervals depending on the quantity of inhomogeneities in the
resonator. We carried out the modeling of semiconductor laser based
on QB with random inhomogeneities. For the purpose of this we used a
quasi-optical millimeter wave cavity resonator filled with random
inhomogeneities and with an inserted Gunn diode as an active
element. In such a self-oscillation system we studied
self-excitation oscillation conditions.  We found out that the
generation is unstable and multi-frequency exists in an empty
resonator near the excitation threshold because of dense spectrum.
Such generation can be explained by frequency jumps in dense
resonator spectrum. These jumps disappear because of "rarefaction"
of the spectrum if inhomogeneities are inserted into the resonator,
and total number of generating frequencies is decreased at a
definite range of control parameters deviation.

\section{\label{sec:level} Statistical spectral theory of the resonator with random bulk inhomogeneities. }

\subsection{\label{sec:level2_11}Statement of the problem}

\begin{figure}
\includegraphics[width=.9\columnwidth]{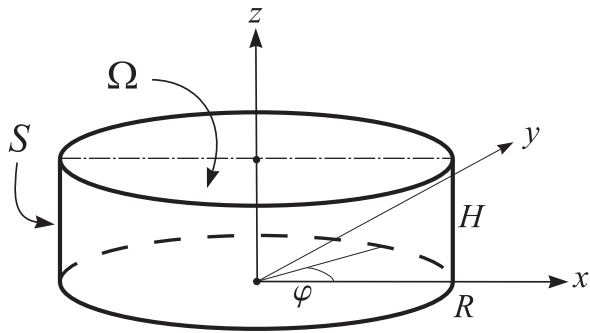}\vspace{1em}
\caption{\label{fig:epsart} The geometry of the cylindrical
quasi-optical cavity resonator. $S$ is the resonator side face,
$\Omega $ is its volume, $H$ is the height of the cylinder, $R$ is
the radius of its base. \label{fig1}}
\end{figure}

Let us consider a cylindrical quasi-optical cavity resonator of
radius $R$  and height $H$ (see Fig.\ref{fig1}). Inner volume of the
resonator, $\Omega $, is assumed to be filled with the material
having random inhomogeneous permittivity. We will be interested in
oscillations that turn into transverse-electrical resonance mode
($TE$-mode) provided that the resonator is filled homogeneously. The
vertical ($z$) component of the electrical field of this mode is
equal to zero.

According to Ref.~\cite{20}, electromagnetic field of the $TE$ mode
can be calculated trough magnetic Hertz vector having only one
non-zero component, namely, {$z$-component} $\Pi _z ({\bf{r}})$. We
assume that inhomogeneity of permittivity of the resonator infill is
small. In this case to define $\Pi _z ({\bf{r}})$ in the
inhomogeneous resonator we can use the approximate wave equation
\begin{equation}\label{Eq1}
\left[ {\Delta  + k^2 \varepsilon ({\bf{r}})} \right]\Pi _z
({\bf{r}}) = 0\ ,
\end{equation}
where all components of the Hertz vector, except z-component,  are
assumed of zero value taking into account small inhomogeneity. In
Eq.(\ref{Eq1}), $\Delta $ is the three-dimensional (3D) Laplacian, $
\varepsilon ({\bf{r}}) = \varepsilon _0  + \delta \varepsilon
({\bf{r}}) + i\alpha $ is the complex permittivity whose imaginary
part $\alpha$ takes phenomenologically into account ohmic loss in
the system; the function $\delta \varepsilon ({\bf{r}})$ describes
random space fluctuations of the permittivity around its average
value $ \varepsilon _0 $, $k = \omega /c$ is wavenumber.

In this paper we are interested in the analogy between
electromagnetic resonators and solid-state micro-objects, namely,
QDs and QBs. In the case of classical resonance system, the problem
of excitation by a given point monochromatic source is governed by
the equation Eq.(\ref{Eq1}) which should be supplied with
$\delta$-term in the right-hand side. The equation thus obtained
coincides in form with the equation for Green function of quantum
particles moving in dissipative medium and being subjected to
inhomogeneous scalar potential. Having this in mind we will carry
out all further analysis relevant to the electromagnetic resonator
in terms of the dynamic equation for particles in the quantum dot of
cylindrical shape.

The equation for Green function of quantum particles moving in
dissipative medium under the influence of random static potential
$V({\bf{r}})$ has the form
\begin{equation}\label{Eq2}
\left[ {\Delta  + k^2  - i/\tau _d  - V({\bf{r}})}
\right]G({\bf{r}},{\bf{r}}') = \delta ({\bf{r}} - {\bf{r}}')\ .
\end{equation}
Here $\tau _d $ is the dissipative attenuation time whose inverse
value has the same physical meaning as the imaginary part of
function $\varepsilon ({\bf{r}})$ in Eq.\eqref{Eq1} taken with minus
sign. The potential $V({\bf{r}})$ in the case of electromagnetic
system is given by $ V({\bf{r}}) = - k^2 \delta \varepsilon
({\bf{r}})$. Boundary conditions for the solution to Eq.(\ref{Eq2})
result from the requirement of vanishing tangential components of
the electrical field of \emph{TE} mode on the resonator interface.
At side boundary $S$, the Neuman condition
\begin{subequations}
\label{eq:whole_1}
\begin{equation}
\left. {\frac{{\partial G({\bf{r}},{\bf{r}}')}}{{\partial r}}}
\right|_{\,S}  = 0 , \label{subeq:1}
\end{equation}
should be met whereas at end surfaces $z=\pm H/2$ the Dirichlet
condition
\begin{equation}
 {G({\bf{r}},{\bf{r}}')} \Big|_{z =  \pm H/2}  = 0
\label{subeq:2}
\end{equation}
\end{subequations}
should be satisfied.

For studying oscillations spectrum of the resonator with random
inhomogeneities, the poles of Green function averaged over
realizations of the potential $V({\bf{r}})$ from Eq.(\ref{Eq2})
should be determined. This function can be found, for example, from
Dyson equation. Yet there by now do not exist effectual methods for
solving this equation in the case of confined multi-dimensional
systems. In this paper, we apply for this purpose the original
calculation technique which relies on precise separation of modes in
an arbitrary confined system, including the disordered one. The
method of mode separation was previously developed for solving
transport problems in disordered 2D open systems \cite{18} and then
modified for systems of waveguide geometry in three dimensions
\cite{19,21,22}. Below we set forth adaptation of this technique for
systems of closed geometry, in particular, for cavity resonators and
quantum dots.

Let us at first turn to mode representation of the equation
Eq.(\ref{Eq2}) using some set of basis functions. The most
appropriate for our purposes seems to be the whole set of
eigenfunctions of the Laplace operator. For the cylindrical
resonator shown in Fig.\ref{fig1} these functions can be factorized
to the form
\begin{equation}\label{Eq4}
\left| {{\bf{r}},{\bm{\mu }}} \right\rangle  = \left| {r,\varphi
;l,n} \right\rangle \left| {z,q} \right\rangle\ ,
\end{equation}
where ${\bf{r}} = (r,\varphi ,z)$ is the radius-vector  in
cylindrical coordinates, ${\bm{\mu }} = \left( {l,n,q} \right)$ is
the vectorial mode index conjugate to that vector. Normalized
eigenfunctions of ``transverse'' part of the Laplacian, which obey
boundary conditions Eq.(\ref{subeq:1}), are given by
\begin{subequations}
\label{eq:whole_2}
\begin{align}\label{transLapleig}
 &\left| {r,\varphi ; l,n} \right\rangle = C_{ln} /\left( {\sqrt \pi  R}
 \right)J_{|n|} \left( {\gamma _l^{(|n|)} r/R} \right)e^{in\varphi }
 \\[3pt]
 &\hspace{1ex}l = 1,2, \ldots ,\quad n = 0, \pm 1, \pm 2, \ldots\
 ,\notag\ ,
\end{align}
where the coefficient $ C_{ln} $ has the form
\begin{equation}
C_{ln}  = \frac{{\gamma _l^{(|n|)} }}{{\left[ {\left( {\gamma
_l^{(|n|)} } \right)^2  - n^2 } \right]^{1/2} \,J_{|n|} \left(
{\gamma _l^{(|n|)} } \right)}}\ .
\end{equation}
\end{subequations}
The set of coefficients $ \gamma _l^{(|n|)} $  in
Eqs.(\ref{eq:whole_2}) consists of positive zeros of the function
$J'_{|n|} \left( t \right)$ which are numbered by index $l$ in
ascending order. The eigenvalues corresponding to functions
Eq.(\ref{transLapleig}) are equal to $\xi _{ln}  = - \left(
{{{\gamma _l^{(|n|)} } \mathord{\left/
 {\vphantom {{\gamma _l^{(|n|)} } R}} \right.
 \kern-\nulldelimiterspace} R}} \right)^2 $.
Basis functions of the ``longitudinal'' Laplacian, ${{\partial
^{{\kern 1pt} 2} } \mathord{\left/
 {\vphantom {{\partial ^{{\kern 1pt} 2} } {\partial {\kern 1pt} z^2 }}} \right.
 \kern-\nulldelimiterspace} {\partial {\kern 1pt} z^2 }}$,
which meet boundary conditions Eq.(\ref{subeq:2}),  are given by
\begin{align}\label{longLapl}
 & \left| {z;q} \right\rangle = \sqrt {\frac{2}{H}}
 \sin \left[ {\left( {\frac{z}{H} + \frac{1}{2}} \right)\pi q} \right]
 \\[3mm]
 & q = 1,2,\; \ldots \;\ ,\notag
\end{align}
the corresponding eigenvalues being equal to $\left( {\pi q/H}
\right)^2 $.

In the basis of functions \eqref{Eq4}, Eq.\eqref{Eq2} takes the form
\begin{equation}\label{Eq7}
\left( {k^2  - \kappa _{\bm{\mu }}^2  - i/\tau _d  - {\cal
V}_{\bm{\mu }} } \right)G_{{\bm{\mu} \bm{\mu }}'}  -
\sum\limits_{{\bm{\nu }} \ne {\bm{\mu }}}^{} {{\cal U}_{{\bm{\mu
\bm{\nu} }}} } G_{{\bm{\nu \bm{\mu} }}'}  = \delta _{{\bm{\mu
\bm{\mu} }}'}\ .
\end{equation}
Here, $G_{{\bm{\mu \bm{\mu}' }}} $ is the Green function in mode
representation, the parameter
\begin{equation}
\kappa _{\bm{\mu }}^2  = \left( {\frac{{\gamma _l^{(|n|)} }}{R}}
\right)^2  + \left( {\frac{{\pi q}}{H}} \right)^2
\end{equation}
is the unperturbed  ``energy'' of the mode ${\bm{\mu }}$ (the
eigenvalue of 3D Laplace operator), functions $ {\cal U}_{{\bm{\mu
\bm{\nu} }}} $ are mode matrix elements of the random potential,
\begin{equation}
{\cal U}_{{\bm{\mu \bm{\nu} }}}  = \int_\Omega  {d{\bf{r}}}
\left\langle {{\bf{r}};{\bm{\mu }}} \right|V({\bf{r}})\left|
{{\bf{r}};{\bm{\nu }}} \right\rangle
\end{equation}

Attention should be drawn to the fact that the \emph{intramode},
i.\,e. diagonal in mode indices, matrix element $ {\cal U}_{{\bm{\mu
\bm{\mu} }}}  \equiv {\cal V}_{\bm{\mu }} $ is separated in
Eq.(\ref{Eq7}) from other terms of the sum where thus only matrix
elements corresponding to \emph{intermode} scattering are left. It
was shown in Ref.~\cite{18} that such a separation of intra- and
intermode effective potentials provides mathematical correctness of
the derivation of closed equations for the diagonal components of
Green matrix $\left\| {G_{{\bm{\mu \bm{\mu} }}'} } \right\|
 $ and through those components for that matrix integrally.

\subsection{\label{sec:level2_12}Separation of the modes}

To solve the infinite set of coupled equations Eq.(\ref{Eq7}) is not
less intricate problem than direct solution of multi-dimensional
differential equation Eq.(\ref{Eq2}). The problem would be resolved
elementary if the resonator modes allowed for their strict
separation. Normally, modes are easily separable if the resonator
has no inhomogeneities. It will be shown below, based on the
technics developed in Refs.~\cite{18, 19, 21}, that in fact they can
be separated even in the case of arbitrarily inhomogeneous
resonator. But the cost of this separation in the general case is
the appearance in equations for each of the modes of the effective
potential, known as \emph{T}-matrix in quantum theory of scattering
\cite{23}, whose functional structure is much more involved than
that of the initial potential $V(\mathbf{r})$.

As a starting point for mode separation we introduce unperturbed (or
trial) mode propagator $G_{\bm{\nu }}^{(V)} $ by omitting in
Eq.(\ref{Eq7}) all intermode potentials $ {\cal U}_{{\bm{\mu \nu }}}
$,
\begin{equation}\label{Eq10}
G_{\bm{\nu }}^{(V)}  = \left( {k^2  - \kappa _{\bm{\nu }}^2  -
i/\tau _d  - {\cal V}_{\bm{\nu }} } \right)^{ - 1}
\end{equation}
The term ``unperturbed'' will thus be used hereupon with respect to
intermode potentials, intramode ones being accounted for precisely.

By substituting ${\bm{\mu '}} = {\bm{\mu }}$ in Eq.(\ref{Eq7}) we
obtain linear non-uniform connection of intramode propagator
$G_{{\bm{\mu \mu }}} $ with all intermode Green functions having the
particular right-hand mode index ${\bm{\mu }}$,
\begin{equation}\label{Eq11}
G_{{\bm{\mu \mu }}}  = G_{\bm{\mu }}^{(V)} \left( {1 +
\sum\limits_{{\bm{\nu }} \ne {\bm{\mu }}} {{\cal U}_{{\bm{\mu \nu
}}} } G_{{\bm{\nu \mu }}} } \right)\ .
\end{equation}
Assuming then ${\bm{\mu }}' \ne {\bm{\mu }}$ and performing some
necessary re-labellings of mode indices we can reduce Eq.(\ref{Eq7})
to the form
\begin{align}\label{Eq12}
 & \left[ {G_{\bm{\nu }}^{(V)} } \right]^{ - 1} G_{{\bm{\nu \mu }}}  -
 \sum\limits_{\scriptstyle {\bm{\nu }}' \ne {\bm{\nu }} \hfill \atop
  \scriptstyle {\bm{\nu }}' \ne {\bm{\mu }} \hfill} {{\cal U}_{{\bm{\nu \nu }}'} }
  G_{{\bm{\nu }}'{\bm{\mu }}}  = {\cal U}_{{\bm{\nu \mu }}} G_{{\bm{\mu \mu }}}  \\
 & ({\bm{\nu }} \ne {\bm{\mu }}) \ .\notag
\end{align}
The latter system of interconnected equations can be solved with
respect to intermode elements of the Green matrix. In the operator
form the solution is given by
\begin{equation}\label{Eq13}
G_{{\bm{\nu \mu }}}  = {\bf{\hat P}}_{\bm{\nu }} \left( {1 - \hat R}
\right)^{ - 1} \hat R{\bf{\hat P}}_{\bm{\mu }} G_{{\bm{\mu \mu }}}\
,
\end{equation}
where the linear operator $\hat R = \hat G^{(V)} \hat {\cal U}$  of
intermode scattering is introduced, which acts in mode subspace
$\overline M _{\bm{\mu }} $ consisting of the whole set of mode
indices but the index ${\bm{\mu }}$; ${\bf{\hat P}}_{\bm{\mu }} $ is
the projection operator whose action reduces to the assignment of
the value  ${\bm{\mu }}$ to the nearest mode index of any adjacent
operator, no matter standing to the left or to the right of
${\bf{\hat P}}_{\bm{\mu }} $. Operators $\hat G^{(V)} $ and $\hat
{\cal U}$ are specified on $\overline M _{\bm{\mu }} $ by matrix
elements
\begin{subequations}
\label{eq:whole_3}
\begin{align}
& \left\langle {\bm{\nu }} \right|\hat G^{(V)} \left| {{\bm{\nu }}'}
\right\rangle  = G_{\bm{\nu }}^{(V)} \delta _{{\bm{\nu \nu }}'}\ ,\\
& \left\langle {\bm{\nu }} \right|\hat {\cal U}\left| {{\bm{\nu }}'}
\right\rangle  = {\cal U}_{{\bm{\nu \nu }}'}\ .
\end{align}
\end{subequations}
Correspondingly, matrix elements of the operator $\hat R$ are given
by
\begin{equation}\label{Eq15}
\left\langle {\bm{\nu }} \right|\hat R\left| {{\bm{\nu }}'}
\right\rangle  = G_{\bm{\nu }}^{(V)} {\cal U}_{{\bm{\nu \nu }}'}\ .
\end{equation}

By substituting intermode propagators in the form Eq.(\ref{Eq13})
into the relationship Eq.(\ref{Eq11}) we obtain the following
rigorous expression for intramode propagator~$ G_{{\bm{\mu \mu }}}$,
\begin{equation}\label{Eq16}
G_{{\bm{\mu \mu }}}  = \left( {k^2  - \kappa _{\bm{\mu }}^2  -
i/\tau _d  - {\cal V}_{\bm{\mu }}  - {\cal T}_{\bm{\mu }} }
\right)^{ - 1}\ .
\end{equation}
Here
\begin{equation}\label{Eq17}
{\cal T}_{\bm{\mu }}  = {\bf{\hat P}}_{\bm{\mu }} \hat {\cal
U}\left( {1 - \hat R} \right)^{ - 1} \hat R{\kern 1pt} {\bf{\hat
P}}_{\bm{\mu }}
\end{equation}
is the portion of the mode ${\bm{\mu }}$ eigen-energy which relates
to the intermode scattering.

It should be noted that for determining the disordered resonator
spectrum the poles of solely diagonal elements of the Green matrix
suffice to be found. Just these elements determine, in view of
relationship Eq.(\ref{Eq13}), all major analytical properties of the
whole Green function from Eq.(\ref{Eq2}). Below we will analyze the
cavity-resonator spectrum in frames of relatively simple statistical
model of random potential $V(\mathbf{r})$.

\subsection{\label{sec:level2_13}Statistical analysis of the resonator spectrum}

We assume the potential $V(\mathbf{r})$ to have zero mean value,
$\left\langle {V(r)} \right\rangle  = 0$ , and the binary
correlation function
\begin{equation}\label{Eq18}
\left\langle {V({\bf{r}})V({\bf{r}}')} \right\rangle  = DW({\bf{r}}
- {\bf{r}}')\ .
\end{equation}
Bearing in mind the forthcoming numerical analysis we will take the
function $W({\bf{r}})$ in the form of Gaussian exponent, viz.
$W({\bf{r}}) = \exp ( - {\bf{r}}^2 /2r_c^2 )$ , where $r_c^{} $
stands for the correlation radius. In the case of electromagnetic
resonator the normalization constant $ D $ in Eq.(\ref{Eq18}) is
given by $D = k^4 \sigma ^2 $, where $\sigma ^2 = \left\langle
{\delta \varepsilon ^2 ({\bf{r}})} \right\rangle $ is the variance
of permittivity fluctuations.

The pair of selected statistical parameters, namely, the average
random potential and its binary correlation function, are sufficient
for making detailed analysis of the system in study if the function
$\delta \varepsilon ({\bf{r}})$ is the Gaussian-distributed random
variable. Yet these two parameters suffice for carrying out the
asymptotically correct analysis even in the case where statistics of
the fluctuations is markedly non-Gaussian, provided that the
potential $ V({\bf{r}}) $ is in a certain sense small. As it is
conventional in condensed matter physics, we will regard the
potential to be small and the resulting scattering, correspondingly,
weak provided that the scattering rate calculated in Born
approximation is small as compared to the unperturbed quasiparticle
energy, $k^2 $ in our case. The smallness of one-fold scattering
probability enables one to regard the potential $V({\bf{r}})$ with
parametric accuracy as Gaussian random process, whatever in fact its
distribution is \cite{24}.

Consider now the self-energy operator of the mode $\bm{\mu}$ which
consists, in accord with Eq.(\ref{Eq16}), of two terms,
\begin{equation}\label{Eq19}
  \Sigma _{\bm{\mu}} = V_{\bm{\mu}}  + {\cal T}_{\bm{\mu}}\ .
\end{equation}
The first term in the right-hand side of this formula vanishes when
being averaged whereas the second term does not. Its average value
can be easily calculated under the presumption of weak scattering.
The strength of the intermode scattering is estimated by the norm of
the operator $\hat R$ entering Eq.(\ref{Eq17}). Assuming this norm
to be small as compared to unity and keeping only two main terms in
the expansion of the inverse operator in Eq.(\ref{Eq17}) we come to
the necessity of performing the averaging not of the exact operator
potential ${\cal T}_{\bm{\mu }}$ but rather of its relatively
compact limiting value
\begin{equation}\label{Eq20}
{\cal T}_{\bm{\mu }}  \approx {\bf{\hat P}}_{\bm{\mu }} \hat {\cal
U}\hat G^{(V)} \hat {\cal U}{\bf{\hat P}}_{\bm{\mu }}  =
\sum\limits_{{\bm{\nu }} \ne {\bm{\mu }}} {{\cal U}_{{\bm{\mu \nu
}}} } G_{\bm{\nu }}^{(V)} {\cal U}_{{\bm{\nu \mu }}}\ .
\end{equation}

When calculating the quantity $\big< {{\cal T}_{\bm{\mu }} } \big>$,
one can neglect with parametric accuracy the correlation between
intramode and intermode potentials. This enables us to average
intramode propagator $G_\nu ^{(V)} $ and its envelopes consisting of
intermode potentials ${\cal U}_{{\bm{\mu \nu }}}$ independently. To
average the function $G_\nu ^{(V)} $ it is worthwhile to present it,
at first, in the integral form,
\begin{equation}\label{Eq21}
G_{\bm{\nu }}^{(V)}  = \int\limits_0^\infty  {dt\,\exp \left[ { -
i\left( {k^2  - \kappa _{\bm{\nu }}^2  - i/\tau _d  - {\cal
V}_{\bm{\nu }} } \right)t} \right]}\ .
\end{equation}
Then the averaging of the integrand in Eq.(\ref{Eq21}) with the aid
of continual integration with Gaussian functional weight yields
\begin{widetext}
\begin{eqnarray}
 \langle G_{\bm{\nu }}^{(V)} \rangle & = & \frac{1}{{k^2
 }}\int\limits_0^\infty  {dt{\kern 1pt} {\kern 1pt} \exp \left[ { -
 i\left( {1 - {{\kappa _{\bm{\nu }}^2 } \mathord{\left/ {\vphantom
 {{\kappa _{\bm{\nu }}^2 } {k^2 }}} \right.
 \kern-\nulldelimiterspace} {k^2 }} - {i \mathord{\left/ {\vphantom
 {i {k^2 \tau _d }}} \right. \kern-\nulldelimiterspace} {k^2 \tau _d
 }}} \right)t - \frac{{t^2 }}{2}\sigma ^2 L_{\bm{\nu }} (r_c )}
 \right]  } \nonumber \\
 & = & \frac{1}{{k^2 }}\sqrt {\frac{\pi
 }{{2\sigma ^2 L_{\bm{\nu }} (r_c )}}} \exp \left[ { - \frac{{\left(
 {k^2  - \kappa _{\bm{\nu }}^2  - {i \mathord{\left/ {\vphantom {i
 {\tau _d }}} \right.\kern-\nulldelimiterspace} {\tau _d
 }}} \right)^2 }}{{2k^4 \sigma ^2 L_{\bm{\nu }} (r_c )}}}
 \right]\left\{ {1 - \Phi \left[ {\frac{{i\left( {k^2  - \kappa
 _{\bm{\nu }}^2  - {i \mathord{\left/ {\vphantom {i {\tau _d }}}
 \right. \kern-\nulldelimiterspace} {\tau _d }}}
 \right)}}{{\sqrt {2k^4 \sigma ^2 L_{\bm{\nu }} (r_c )} }}} \right]}
 \right\}\ .
\label{Eq22}
\end{eqnarray}
\end{widetext}
Here $\Phi (\xi )$ is the probability integral \cite{25}, $L_\nu
(r_c )$ is the dimensionless correlator of intramode potential
${\cal V}_{\bm{\nu }}$. In the case of Gaussian correlation function
$W({\bf{r}})$ we readily obtain
\begin{widetext}
\begin{eqnarray}
 L_{\bm{\nu }} (r_c ) & =& \frac{1}{{k^4 }}\left\langle {{\cal V}_{\bm{\nu }} {\cal V}_{\bm{\nu }} }
 \right\rangle  \nonumber
 = \iint\limits_{\Omega}{d{\bf{r}}{\kern 1pt} d{\bf{r}}'\left\langle {{\bf{r}},{\bf{\nu }}}
\right.\left| {{\bf{r}},{\bf{\nu }}} \right\rangle }
 \exp \left[ { - {{({\bm{r}} - {\bm{r}}')^2 } \mathord{\left/
 {\vphantom {{({\bf{r}} - {\bf{r}}')^2 } {2r_c^2 }}} \right.
 \kern-\nulldelimiterspace} {2r_c^2 }}} \right]\left\langle {{\bf{r}}',{\bm{\nu }}} \right.
 \left| {{\bf{r}}',{\bm{\nu }}} \right\rangle
 \\ \nonumber
  &=& \frac{8}{\pi }C^4_{l_{\bm{\nu }} n_{\bm{\nu }} } \int\limits_0^1 {\int\limits_0^1 {dsds'
  \exp \left[ { - \frac{{H^2 }}{{2r_c^2 }}(s - s')^2 } \right]} } \sin ^2 (\pi q_{n_{\bm{\nu }} } s)
  \sin ^2 (\pi q_{n_{\bm{\nu }} } s') \\
  &&\times \int\limits_0^1 {\int\limits_0^1 {tt'dtdt'
  J_{|n_{_{\bm{\nu }} } |}^2 } } (\gamma _{l_{\bm{\nu }}^{} }^{|n_{\bm{\nu }} |} t)
  J_{|n_{_{\bm{\nu }} } |}^2 (\gamma _{l_{\bm{\nu }}^{} }^{|n_{\bm{\nu }} |} t')
  \oint {d\varphi \exp \left[ { - \frac{{R^2 }}{{2r_c^2 }}\left( {t^2  + t'^2  - 2tt'\cos \varphi } \right)} \right]}\ .
\label{Eq23}
\end{eqnarray}
\end{widetext}

Although the expression Eq.(\ref{Eq22}) is formally exact, it is not
quite convenient for the analysis in view of its bulky structure.
Asymptotical calculations at large and small values of the
probability integral argument permit us to make use of much simpler
interpolation expression,
\begin{equation}\label{Eq24}
\big<G_{\bm{\nu }}^{(V)} \big> \approx \left( {k^2  - \kappa
_{\bm{\nu }}^2  - i/\tau _{\bm{\nu }}^* } \right)^{ - 1}\ ,
\end{equation}
which is close in form to the initial unperturbed Green function
Eq.(\ref{Eq10}). Here we have used the notation ${1/\tau _\nu ^*  =
1/\tau _d^{}  + k^2 \sqrt {(2/\pi )\sigma ^2 L_\nu (r_c )}}$ for the
effective scattering frequency. The latter includes both the initial
dissipative term $1/\tau _d$ and the addendum originating from wave
scattering by random inhomogeneities.

Interpolation of the average trial Green function by the expression
Eq.(\ref{Eq24}) permits us to interpret the quantity $k^2 \sqrt
{(2/\pi )\sigma ^2 L_\nu ^{} (r_c )} $ as the dephasing frequency of
the mode state $\bm{\nu}$, which is related to scattering by the
inhomogeneities in the resonator.

To make further comparison of theoretical results with experimental
data we will consider below the specific limiting case where the
inequalities are fulfilled
\begin{equation}
r_c  \lesssim H \ll R\ .
\end{equation}
These restrictions enable us to calculate the integrals in
Eq.(\ref{Eq23}) asymptotically, thus resulting in the following
estimate for the parameter $L_\nu  (r_c )$: ${L_{\bm{\nu }} (r_c )
\sim \left( {{{r_c } \mathord{\left/ {\vphantom {{r_c } R}} \right.
\kern-\nulldelimiterspace} R}} \right)^2 }$. The correlator of
intermode potentials in Eq.(\ref{Eq20}) can be represented as
\begin{widetext}
\begin{eqnarray}
 \big<{U_{{\bm{\mu \nu }}} U_{{\bm{\nu \mu }}} }\big>
 &=& k^4 \sigma ^2 \left( {\frac{2}{\pi }} \right)^2 \int\limits_0^1
 {\int\limits_0^1 {dsds'} \exp \left[ { - \frac{{H^2 (s - s')^2
 }}{{2r_c ^2 }}} \right]\sin (\pi q_{n_{\bm{\mu }} } s)} \sin (\pi
 q_{n_{\bm{\mu }} } s')\sin (\pi q_{n_{\bm{\nu }} } s)\sin (\pi
 q_{n_{\bm{\nu }} } s') \nonumber \\
 &&\times \int\limits_0^1
 {\int\limits_0^1 {tt'dtdt'J_{\left| {n_{\bm{\mu }} } \right|}
 (\gamma _{l_{\bm{\mu }} }^{(\left| {n_{\bm{\mu }} } \right|)} t)} }
 J_{\left| {n_{\bm{\mu }} } \right|} (\gamma _{l_{\bm{\mu }}
 }^{(\left| {n_{\bm{\mu }} } \right|)} t')J_{\left| {n_{\bm{\nu }} }
 \right|} (\gamma _{l_{\bm{\nu }} }^{(\left| {n_{\bm{\mu }} }
 \right|)} t)J_{\left| {n_{\bm{\nu }} } \right|}
 (\gamma _{l_{\bm{\nu }} }^{(\left| {n_{\bm{\mu }} } \right|)}t')\nonumber
 \\[3mm]
 &&\times \oint\oint {d\varphi d\varphi '\exp \left\{ { -
 i(n_{\bm{\mu }}  - n_{\bm{\nu }} )(\varphi  - \varphi ') -
 \frac{{R^2 }}{{2r_c ^2 }}\left[ {t^2  + t'^2  - 2tt'\cos (\varphi  -
 \varphi ')} \right]} \right\}}\ .
\label{Eq26}
\end{eqnarray}
\end{widetext}
Asymptotic calculation of the integrals over $\varphi$ and
$\varphi'$ results in the following formula
\begin{equation}\label{Eq27}
\big< {\cal U_{{\bm{\mu \nu }}}\cal  U_{{\bm{\nu \mu }}} } \big> =
k^4 \sigma ^2 \frac{{r_c }}{R}A_{{\bm{\mu \nu }}} (r_c )\ ,
\end{equation}
where the factor $A_{{\bm{\mu \nu }}} (r_c )$ is given by
\begin{widetext}
\begin{align}
 A_{{\bm{\mu \nu }}} (r_c ) =& \frac{8}{{\sqrt \pi  }}C_{l_{\bm{\mu
 }} n_{\bm{\mu }} }^2 C_{l_{\bm{\nu }} n_{\bm{\nu }} }^2
 \int\limits_0^1 {\int\limits_0^1 {ds{\kern 1pt} ds'\exp \left[ { -
 \frac{{H^2 }}{{2r_c^2 }}(s - s')^2 } \right]} \,} {\kern 1pt}
 \,{\kern 1pt} \sin \left( {\pi q_{n_{\bm{\mu }} } s} \right)\sin
 \left( {\pi q_{n_{\bm{\mu }} } s'} \right)\sin \left( {\pi
 q_{n_{\bm{\nu }} } s} \right)\sin \left( {\pi q_{n_{\bm{\nu }} } s'}
 \right){\kern 1pt} \nonumber \\
 &\times \int\limits_0^1
 \int\limits_0^1 \sqrt {tt'\,} dt{\kern 1pt} dt'\exp \left[ { -
 \frac{{R^2 }}{{2r_c^2 }}\left( {t - t'} \right)^2 - \frac{{r_c^2
 \left( {n_{\bm{\mu }}  - n_{\bm{\nu }} } \right)^2 }}{{4R^2 tt'}}}
 \right]\notag\\[3mm]
 &\times J_{|n_{\bm{\mu }} |}^{} \left( {\gamma _{l_{\bm{\mu }}
 }^{(|n_{\bm{\mu }} |)} t} \right)J_{|n_{\bm{\mu }} |}^{} \left(
 {\gamma _{l_{\bm{\mu }} }^{(|n_{\bm{\mu }} |)} t'}
 \right)J_{|n_{\bm{\nu }} |}^{} \left( {\gamma _{l_{\bm{\nu }}
 }^{(|n_{\bm{\nu }} |)} t} \right)J_{|n_{\bm{\nu }} |}^{} \left(
 {\gamma _{l_{\bm{\nu }}
 }^{(|n_{\bm{\nu }} |)} t'} \right)  \;.
\label{Eq28}
\end{align}
\end{widetext}

With the results \eqref{Eq24} and \eqref{Eq27}, separation of real
and imaginary parts of the average $T$-matrix, $\left\langle {\cal
T_{\bm{\mu }} } \right\rangle = \Delta k_{\bm{\mu }}^2  + i/\tau
_{\bm{\mu }}^{(ch)} $, results in the following expressions for
shifting and broadening the $\bm{\mu}$-th resonant level,
\begin{subequations}
\label{eq:whole_4}
\begin{equation}
\Delta k_{\bm{\mu }}^2  = k^4 \sigma ^2 \frac{{r_c
}}{R}\sum\limits_{{\bm{\nu }} \ne {\bm{\mu }}} {A_{{\bm{\mu \nu }}}
(r_c ){\mathop{\rm Re}\nolimits} \langle {G_{\bm{\nu }}^{(V)} }
\rangle }\ , \label{Eq29a}
\end{equation}
\begin{equation}
\frac{1}{{\tau _{\bm{\mu }}^{(ch)} }} = k^4 \sigma ^2 \frac{{r_c
}}{R}\sum\limits_{{\bm{\nu }} \ne {\bm{\mu }}} {A_{{\bm{\mu \nu }}}
(r_c ){\mathop{\rm Im}\nolimits} \langle {G_{\bm{\nu }}^{(V)} }
\rangle }\ . \label{Eq29b}
\end{equation}
\end{subequations}
The factor $A_{{\bm{\mu \nu }}} (r_c )$, as can be seen from
Eq.(\ref{Eq28}), is a real-valued quantity, its absolute value being
estimated as $r_c /R \ll 1$. Although the sign of this factor cannot
be uniquely identified in the general case since it contains the
dependence on specific indices of modes between which the scattering
is carried out, numerical analysis shows that $ A_{{\bm{\mu \nu }}}
(r_c ) > 0 $, which corresponds to the broadening of spectral lines.
One can adequately estimate both the shift and the broadening of
$\bm\mu$-th resonant level by substituting the function $\big<
G_{\bm{\nu }}^{(V)} \big>$ in the interpolated form Eq.(\ref{Eq24}),
instead of its exact expression Eq.(\ref{Eq22}), into the right-hand
sides of formulas Eq.(\ref{eq:whole_4}). The result in this case
reduces to
\begin{subequations}
\label{eq:whole_5}
\begin{equation}
\Delta \kappa _{\bm{\mu }}^2  = k^4 \sigma ^2 \frac{{r_c
}}{R}\sum\limits_{{\bm{\nu }} \ne {\bm{\mu }}} {A_{{\bm{\mu \nu }}}
(r_c )\frac{{\kappa _{\bm{\mu }}^2  - \kappa _{\bm{\nu }}^2
}}{{\left( {\kappa _{\bm{\mu }} ^2  - \kappa _{\bm{\nu }}^2 }
\right)^2  + \left( {{1 \mathord{\left/ {\vphantom {1 {\tau
_{\bm{\nu }}^* }}} \right. \kern-\nulldelimiterspace} {\tau
_{\bm{\nu }}^* }}} \right)^2 }}}\ , \label{Eq30a}
\end{equation}
\begin{equation}
\frac{1}{{\tau _{\bm{\mu }}^{(ch)} }} = k^4 \sigma ^2 \frac{{r_c
}}{R}\sum\limits_{{\bm{\nu }} \ne {\bm{\mu }}} {A_{{\bm{\mu \nu }}}
(r_c )\frac{{{1 \mathord{\left/ {\vphantom {1 {\tau _{\bm{\nu }}^*
}}} \right. \kern-\nulldelimiterspace} {\tau _{\bm{\nu }}^*
}}}}{{\left( {\kappa _{\bm{\mu }} ^2  - \kappa _{\bm{\nu }}^2 }
\right)^2  + \left( {{1 \mathord{\left/ {\vphantom {1 {\tau
_{\bm{\nu }}^* }}} \right. \kern-\nulldelimiterspace} {\tau
_{\bm{\nu }}^* }}} \right)^2 }}}\ . \label{Eq30b}
\end{equation}
\end{subequations}

The structure of the summands in Eqs.(\ref{eq:whole_5}) indicates
unambiguously that both the shift and the broadening of each given
resonance $(k^2 \simeq k_{\bm{\mu }}^2 )$ are mainly provided by its
interactions with the adjacent resonances, whose position on the
frequency axis are confined to the region limited by order equality
$\left| {\kappa _{\bm{\mu }}^2  - \kappa _{\bm{\nu }}^2 } \right|
\sim 1/\tau _{\bm{\nu }}^ * $. In the case where only one resonance
level with, say, frequency $\overline{\omega}_{\bm{\nu }} $ proves
to fall into the above indicated interval around the ${\bm{\mu
}}$-th resonance (we assume the condition $\sigma r_c /R \ll 1$ of
weak intramode scattering to be met) its contribution to the shift
and the width of the ${\bm{\mu }}$-th resonance is estimated as
\begin{subequations}
\label{eq:whole_6}
\begin{equation}
\omega _{\bm{\mu }}  - \omega _{{\bm{\mu }}0}  = \frac{{\sigma ^2
\lambda A_{{\bm{\mu \nu }}} (\omega _{{\bm{\mu }}0}  -
\overline{\omega} _{\bm{\nu }} )}}{{(\delta \omega _{{\bm{\mu }}0} +
\sigma \lambda )^2 }}\ ,
\end{equation}
\begin{equation}
\delta \omega _{\bm{\mu }}  = \frac{{\sigma ^2 \lambda A_{{\bm{\mu
\nu }}} }}{{\delta \omega _{{\bm{\mu }}0}  + \sigma \lambda }}\ .
\end{equation}
\end{subequations}
Here $ \omega_{{\bm{\mu}}0 }$ and $ \omega_{\bm{\mu}} $ are the
cyclic spectral frequency of empty resonator and that of the
resonator filled with random inhomogeneities, respectively, $\delta
\omega _{{\bm{\mu }}0} $ and $\delta \omega _{\bm{\mu }} $ are
relative widths of their spectral lines, the parameter ${\lambda
\sim r_c /R}$. If several resonances fall into the indicated
vicinity of the ${\bm{\mu }}$-th resonance simultaneously, they
contribute additively to both the level position and width.

The above presented theory shows that the filling of a quasioptical
cavity resonator with randomly distributed inhomogeneities results
in both the random shift of its resonance lines and the increase of
their linewidths. It should be noted that formulas
Eqs.(\ref{eq:whole_4}) and \eqref{eq:whole_5} were obtained with the
use of some procedure of averaging over the ensemble of
macroscopically identical resonance systems whose microscopic
realizations are different. What this formally means is that the
statistical theory can give only qualitative predictions for the
experiment. The situation seems to be quite similar to that one
faces when studying problems of wave scattering by randomly rough
surfaces \cite{2}. Yet the problem of oscillations in the
quasi-optical cavity resonator filled with bulk random
inhomogeneities differs substantially from the problem of wave
scattering by randomly rough surface. The point is that oscillations
in cavity resonators become established as a result of multiple
transmissions of waves (with multiplicity of the order of $Q$)
through the random infill of the resonator. With this consideration
in mind one might expect that under the restriction $r_c\ll R$ the
characteristics of resonance lines should be effectively
self-averaged \cite{24}. If this is the case, the agreement between
theory and experiment may be achieved either in the course of small
number of trials or even for one particular realization of the
random system. The confirmation or the denial of this assumption may
be achieved through the comparison of theoretical predictions and
the data of corresponding measurements.

\subsection{\label{sec:level2_14}Numerical simulation}

To compare the theory and experiment we calculate the spectrum of
$TE$ modes in the quasi-optical cylinder cavity millimeter wave
resonator filled with random dielectric inhomogeneities using our
theoretical results. At the resonator spectrum calculation the
resonator parameters were taken according to the experiment
conditions.  The spectrum was calculated by solving the excitation
task with pointed external dipole. Fig.2 shows the influence of
random bulk inhomogeneities on the spectrum at the different values
of the parameter $\sigma $ calculated using Eq.(30). As an example
we selected the spectrum in the frequency interval of 36-37 GHz.

\begin{figure}
\includegraphics[width=3.2in]{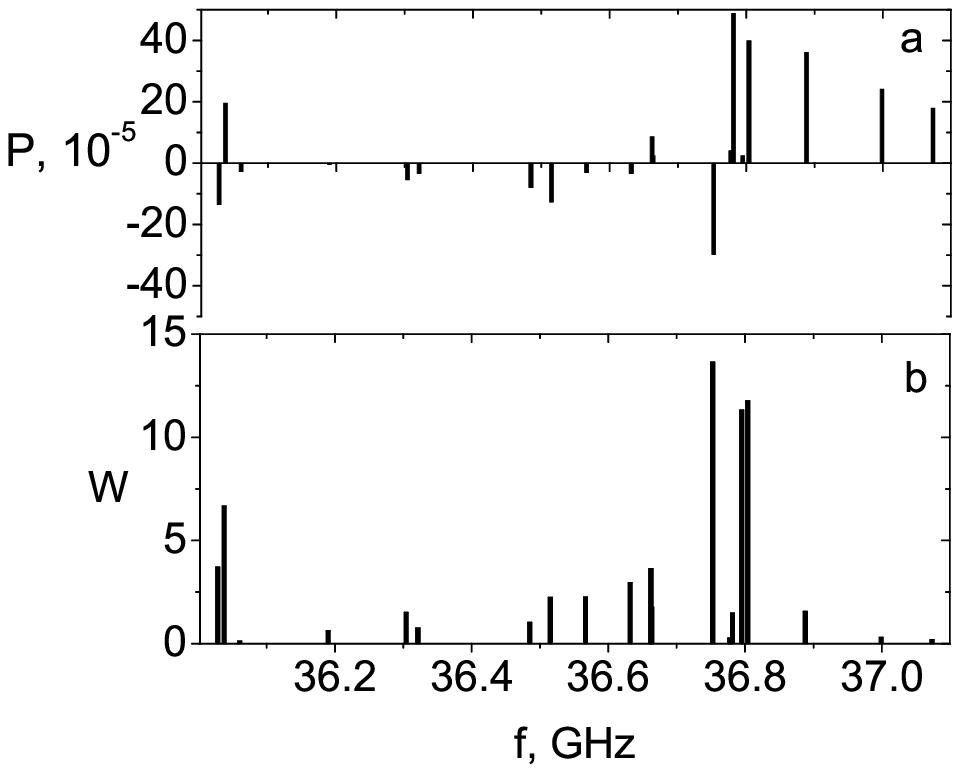}
\includegraphics[width=3.2in]{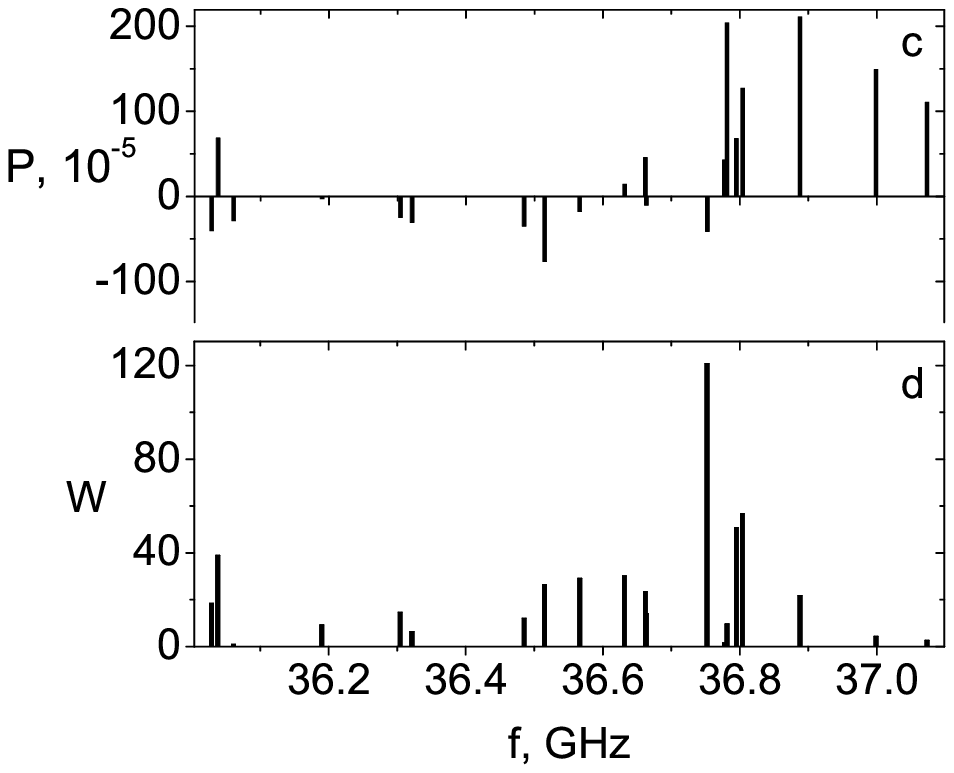}
\caption{\label{fig:epsart} The dependence of relative shifting $P =
\Delta \kappa _{\bm{\mu }}^2 /\kappa _{\bm{\mu }}^2  = \Delta
f_{\bm{\mu }}^2 /f_{\bm{\mu }}^2 $  (a,c) of spectral lines of the
resonator on frequency $f$ caused by bulk inhomogeneities and their
relative broadening $W = \tau _d^{} /\tau _{\bm{\mu }}^{(ch)} =
Q_d^{} /Q_{\bm{\mu }}^{} $ (b,d)  at the permittivity dispersion
values of inhomogeneities: $\sigma  = 0.02$ (a,b) and $\sigma =
0.05$ (c,d); $r_c  = 0.3$ cm. }
\end{figure}

It is shown that the spectrum essentially changed at the presence of
dielectric inhomogeneities in the resonator. The magnitude of
shifting and broadening of spectral lines are caused by resonator
filling by inhomogeneities. The solitary spectral lines even at
presence of inhomogeneities keep high quality factor, and their
broadening is enough small (for example the spectral lines in the
range of 36.2-36.6 GHz).  At the same time the shifting and
broadening of adjacent lines are significant (for example, the
spectral lines near of 36.8 GHz). This results that the number of
excited lines with high quality factor are essentially decreased if
the number of inhomogeneities are getting bigger. The reduction of
the number of high quality lines we can explain as "rarefaction" of
the resonator spectrum caused by unequal scattering conditions of
different modes on inhomogeneities. Fig.3 shows the
amplitude-frequency dependence of the resonator spectrum built as
the frequency dependence of Green function module at different
magnitude of parameter $\sigma $.  So, at the increase of the number
of inhomogeneities in the resonator the adjacent resonances
overlapped and the quality factor of combined resonances is getting
smaller (for example, spectral lines near by 36.8 GHz).

\begin{figure}
\includegraphics[width=3.2in]{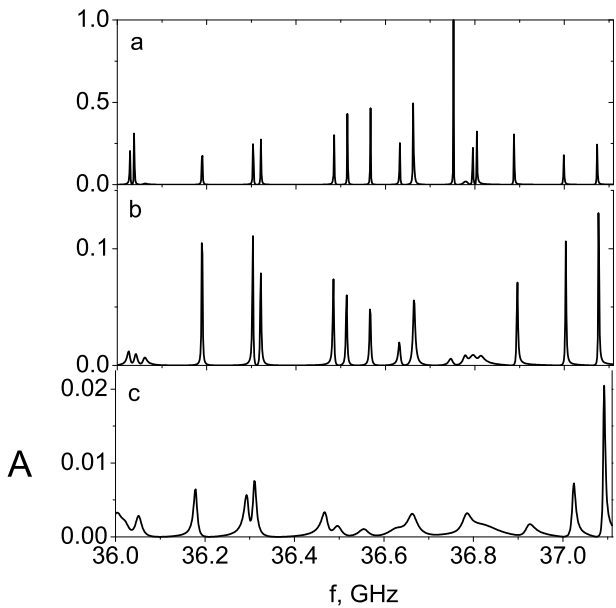}
\caption{\label{fig:epsart} The amplitude-frequency dependence of
the empty resonator spectrum (a) and the spectrum of resonator with
inhomogeneities at $\sigma  = 0.02$ (b) and $\sigma  = 0.05$ (c);
$r_c  = 0.3$ cm. The amplitude is normalized by maximal amplitude of
resonances for the empty resonator in the considered wave range.}
\end{figure}

\section{Experiment and discussion}
The main goal of our experiment is the verification of the numerical
simulation results of shifting and broadening of spectral lines
caused by inhomogeneities and the possibility of resonator spectrum
"rarefaction". The another goal is as follows. In the paper [3] we
detected strong spectrum stochastization caused by inhomogeneities
such as anisotropic sapphire particles with the dimensions of order
of operating wavelength placed into the cavity resonator. The
spectrum has mixed state due to both regular and chaotic spectrum
part exist. In contrast to [3], in the present paper we study the
influence on the resonator spectrum relatively small isotropic
inhomogeneities. Such inhomogeneities can be made by styrofoam
particles with the average value of the permittivity closed to one
and with small dielectric loss angle. Thus, there is a question: is
it possible the spectrum chaotization here?

\subsection{\label{sec:level2_21}Experiment technique}
The study of the influence of inhomogeneities on the cavity
resonator spectrum we carried out at frequency range of 32-37 GHz.
We took a quasi-optical cylinder millimeter wave resonator random
filled with styrofoam particles (Fig.4). The styrofoam particles
have the real part of permittivity is about one, close to the
open-air permittivity and small dielectric loss, $\varepsilon  =
1.04 + i10^{ - 4} $.

\begin{figure}
\includegraphics[width=3.2in]{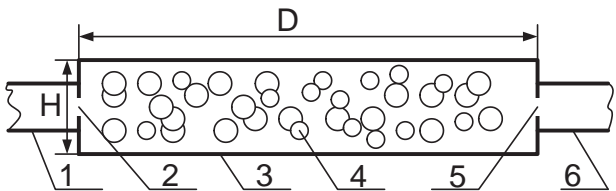}
\caption{\label{fig:epsart} The quasi-optical cylinder cavity
millimeter wave resonator filled with inhomogeneities, 1,6 are
input/output waveguides, 2,5 are holes coupling with waveguides, 3
is  the resonator body, 4 is styrofoam particles, $D$ is the
resonator diameter, $H$ is its hight; $D = 130$ mm, $H =14$ mm.  }
\end{figure}

To study the influence of small dielectric particles on the
resonator spectrum it is necessary to provide high quality factor
for the resonator oscillations without inhomogeneities. For that we
excited $TE$ mode. The magnetic field vector of this mode is
directed along the resonator z-axis, and microwave currents do not
cross the interface between the flat resonator face and cylinder
surface. Owing to that the empty resonator has high quality factor
up to $2 \times 10^4 $ . To excite the selected mode we used a
waveguide diffraction antenna. It is the circular hole with the
diameter of 2 mm in a thin diaphragm with the thickness of 0.1 mm
closing the input waveguide. The diaphragm surface is flush-mounted
with the side-cut cylinder surface of the resonator. The same
antenna is used to receive the oscillations on the opposite side of
the cylinder surface.

The resonator spectrum was detected using "on pass" regime in 32 -
37 GHz wave range  by  the wide frequency  standing wave ratio
meter. Measurement process was automated. Signal from the
measurement device using analog-digital conversion goes into
computer. The further signal processing (the determination of the
spectral line intensity, its quality factor, and frequency) was done
by special software GUI application. It gives resource in a short
space of time to handle measurement data for a huge number of
realizations of the random inhomogeneities distribution in the
resonator. Owing to that the frequency and quality factor
measurement accuracy for spectral lines do not exceed 0.1\% and 1\%,
respectively. The styrofoam particles that are used as
inhomogeneities have the size about of 2-3 mm. The space
distribution of them was arbitrary for each realization.  The
spectral characteristics were measured depending on the number of
these inhomogeneities.

\subsection{\label{sec:level2_2}Shifting and broadening of spectral lines. Effect of spectrum
"rarefaction"}

The spectrum of the empty resonator is dense enough and consists of
84 narrow spectral lines in the range of 32-37 GHz (Fig.5). Each
line was identified according to mode indexes of corresponding to
its eigen resonator oscillation. We found out the oscillations with
high quality factor ($Q$) of order of  $10^4 $ and higher have small
azimuth indexes ($n \sim 1$) and high radial indexes ($l >  > 1$) .
For the base oscillation mode ($n >  > 1,l = 1$) that has the field
distribution concentrated inside of the resonator side-cut
(whispering-gallery oscillations [5]), $Q$ is about $2 \cdot 10^3 $.
The spectrum of the resonator with inhomogeneities is essentially
different from the spectrum of the empty one. By the increase of the
number of inhomogeneities spectral lines are getting wider, and  $Q$
is getting smaller. The intensity of broadening lines is decreased
as well. The spectrum keeps only few lines with enough high
intercity and $Q$ value. For such lines the $Q$ value is close to
the $Q$-factor of the spectral lines in the empty resonator. So, the
number of spectral lines with enough high intensity and with $Q >
10^4 $  is 52 in the empty resonator (Fig.5a), at filling the
resonator by styrofoam particles the number of such lines is 10
(Fig.5b), and at filling the resonator by pressed styrofoam
particles the number of lines is 3 (Fig.5c).

\begin{figure}
\includegraphics[width=3.2in]{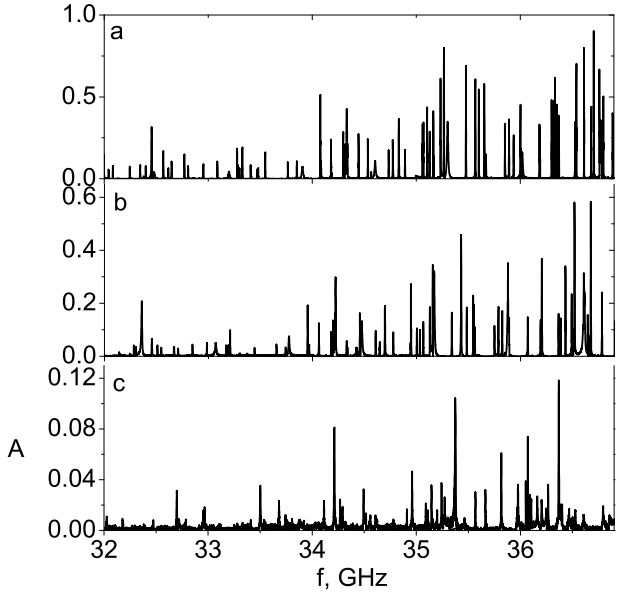}
\caption{\label{fig:epsart} The spectrum of the empty resonator (84
spectral lines) (a); the resonator filled with styrofoam particles
(77 spectral lines) (b); the resonator filled with pressed styrofoam
particles (57 spectral lines) (c); The normalization of amplitude
was made on the maximal amplitude value for the empty resonator.}
\end{figure}

The presence in the resonator with inhomogeneities together with
broadening lines few high $Q$ spectral lines is equivalent to the
spectrum "rarefaction" (Fig.5). It is necessary to note that the
presence of lines with high $Q$ indicates that the broadening is
specified, mainly, by the inhomogeneous of the resonator filling
permittivity and is not specified by additional dissipation loss
caused by small dielectric loss in the styrofoam. Except broadening
of spectral lines they have frequency shifting. This frequency
shifting has both regular component and random one as well (Fig.6).

\begin{figure}
\includegraphics[width=3.2in]{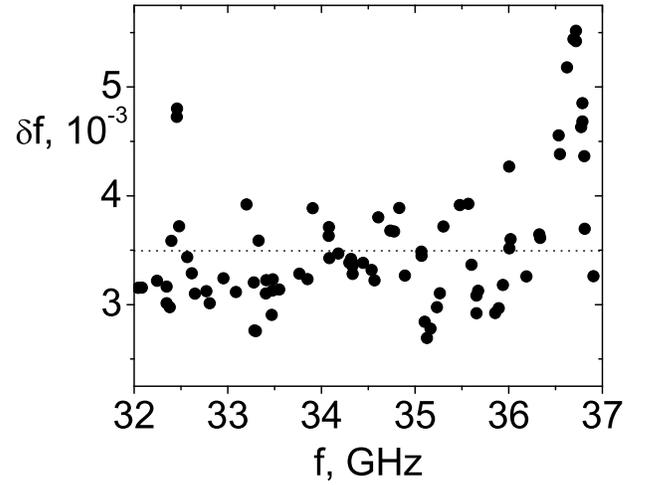}
\caption{\label{fig:epsart} The dependence of the frequency shifting
$\delta f = \left( {f_{empty} - f_{in\hom }} \right)/f_{empty} $ on
frequency $f$ for the resonator filled by inhomogeneities. The
dotted line is the value of shifting regular component.}
\end{figure}

The regular shifting happens towards the low frequency direction
because of the increase of average dielectric permittivity of
inhomogeneous medium in the resonator. For example, it is 150 MGz
for the resonator fully filled with styrofoam particles.

It is significant that if the distance between resonances is enough
small they can both come close and depart from each other because of
influence of inhomogeneities. Their $Q$ is essentially changed
(Fig.7,8): it can increase for one of them and decrease for another.

\begin{figure}
\includegraphics[width=3.2in]{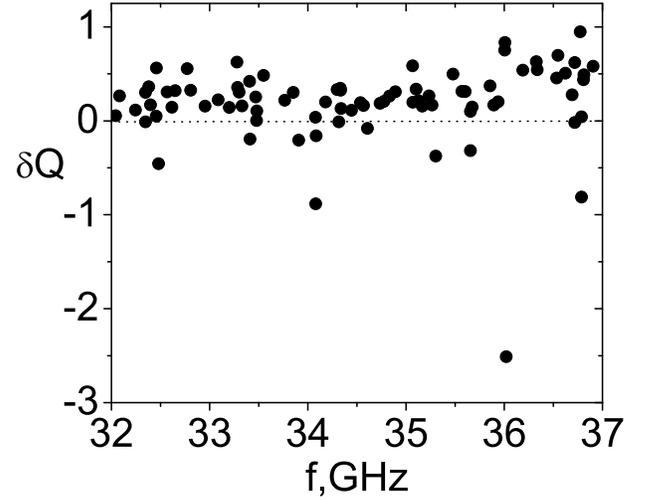}
\caption{\label{fig:epsart} The relative $Q$-factor deviation
$\delta Q = (Q_{ empty }  - Q_{in\hom} )/Q_{empty} $ depends upon
frequency $f$ for the resonator filled by inhomogeneities.  For
major part of resonances the relative $Q$-factor deviation is more
than zero, i.e., their $Q$ is less than for the empty resonator. The
relative $Q$-factor deviation is less than zero for several nearest
adjacent resonances that have overlapped spectral lines (in the case
of the empty resonator) and split into separate resonances at the
presence of inhomogeneities, i.e., the spectral line repulsion takes
place here (see Fig.8). }
\end{figure}

\begin{figure}
\includegraphics[width=3.2in]{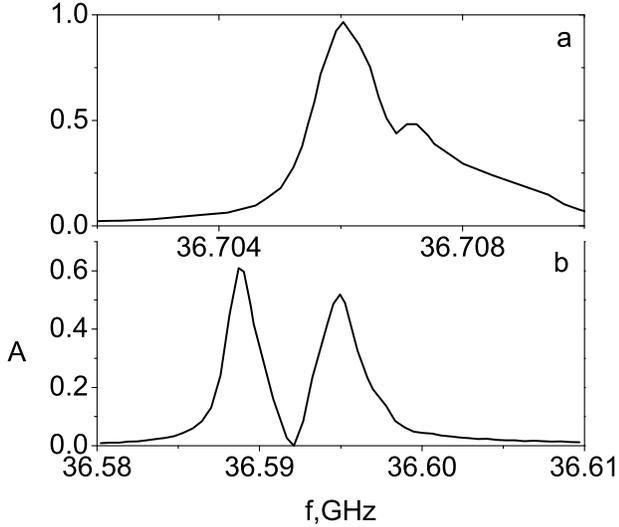}
\caption{\label{fig:epsart}The adjacent resonances with indexes $
{\rm{n = 34, l = 3;}}\;{\rm{n = 4, l = 15;}} $ in the empty
resonator (a) and in the resonator filled by inhomogeneities (b).
The normalization of the spectral lines amplitude was made on the
maximal spectral line amplitude for the empty resonator. }
\end{figure}

Both the spectral lines broadening and their shifting have quality
agreement with described above theory and can be interpreted in the
terms of intermode scattering. The maximal broadening is for the
adjacent lines that is agreed with theory.

We can give the following physical explanation of spectral lines
broadening caused by random inhomogeneities based on quantum
mechanics interpretation of resonator oscillations.   In empty
resonator the lifetime of a resonant mode in a given state
(describing by an eigen resonant frequency) and its spectral line
width define by dissipation loss in the resonator body. The
uncertainty of mode energy within the spectral line width is caused
by this loss. The intermode interaction appears at inserting
inhomogeneities into the resonator. Owing to that the uncertainty of
mode energy and, correspondingly, the spectral line width increase
due to the mode transition between nearest states. The mode energy
loss due to dissipation loss in the resonator body and transition to
the neighboring state are additive to the spectral line width.

\subsection{\label{sec:level2_3}Statistical analysis of IF intervals}
If the number of inhomogeneities increases the spectrum of the
quasi-optical resonator possesses stochastic character that is
visualized in distribution of IF intervals. In order to define the
relation between regular and random spectral components the
comparison of IF intervals distribution obtained experimentally with
different theoretical distributions based on a priori data about
statistical process is usually used. In particular, we use the Brody
function determined the distribution of IF intervals probability
$P_{B} (s)$ that is given by

\begin{equation}
P_B (s) = As^\beta  \exp ( - Bs^{1 + \beta } )
\end{equation}

where $s = (\omega _n  - \omega _{n - 1} )\rho (\omega _n )$,
$\omega _n $ is the spectral line frequency, $\rho (\omega _n^{} )$
is the spectral density - the superposition of regular and random
motion density, $\beta $ is the measure of stochastic motion,
constants $A$ and $B$ are defined from the condition of
standardization: $A = (1 + \beta )B,\quad $, $B = \Gamma ^{1 + \beta
} (2 + \beta )(1 + \beta )^{ - 1} $, $\Gamma (z)$ is the Gamma
function. At $\beta  \to 0$ IF intervals in the spectrum are not
correlated and can be described by the Poisson distribution, and at
$\beta  \to 1$ we have Wigner distribution, when the repulsion
effect of spectral lines exists, that is the probability of closest
to zero inter-frequency interval is equal zero as well.

If stochastization measure in the spectrum is relatively small and
the distribution of IF intervals is close to the Poisson one we can
use Berry-Robnik distribution $P_{BR} (s)$ [26] that is given by

\begin{equation}
\begin{array}{l}
P_{BR} (s) = \rho ^2 e^{ - \rho s} erfc\left( {\frac{{\sqrt \pi  }}{2}\left( {1 - \rho } \right)s} \right) +  \\
\left[ {\frac{\pi }{2}\left( {1 - \rho } \right)^2 s + 2\rho } \right]\left( {1 - \rho } \right)e^{ - \rho s - \frac{\pi }{4}\left( {1 - \rho } \right)^2 s^2 }  \\
\end{array}
\end{equation}

where ${\mathop{\rm erfc}\nolimits} (x) = \frac{2}{{\sqrt \pi
}}\int\limits_x^\infty  {\exp ( - t^2 )dt} $.

$\rho  = 1$ is relative phase volume occupied with regular
trajectory in mixed systems. The limit $\rho \to 1$ corresponds to
regular system; $\rho  \to 0$ is completely chaotic system.  $1 -
\rho$ is relative phase volume occupied with chaotic motion.

The experimental data show that in the empty resonator the IF
intervals distribution is close enough to the Berry-Robnik
distribution with $\rho  = 1$ (Poisson distribution, $P(s) \sim \exp
( - s)$)(Fig.9a). By increase of the number of inhomogeneities
(styrofoam particles) the function $P(s)$ is reduced and it has
maximum at small $s$. The presence of the maximum $P(s)$ indicates
the appearance of the repulsion effect of spectral lines, and the
random component is increased. We found out that at full filling the
resonator by styrofoam particles $\rho  = 0.4$ (Fig.9b).

\begin{figure}
\includegraphics[width=3.2in]{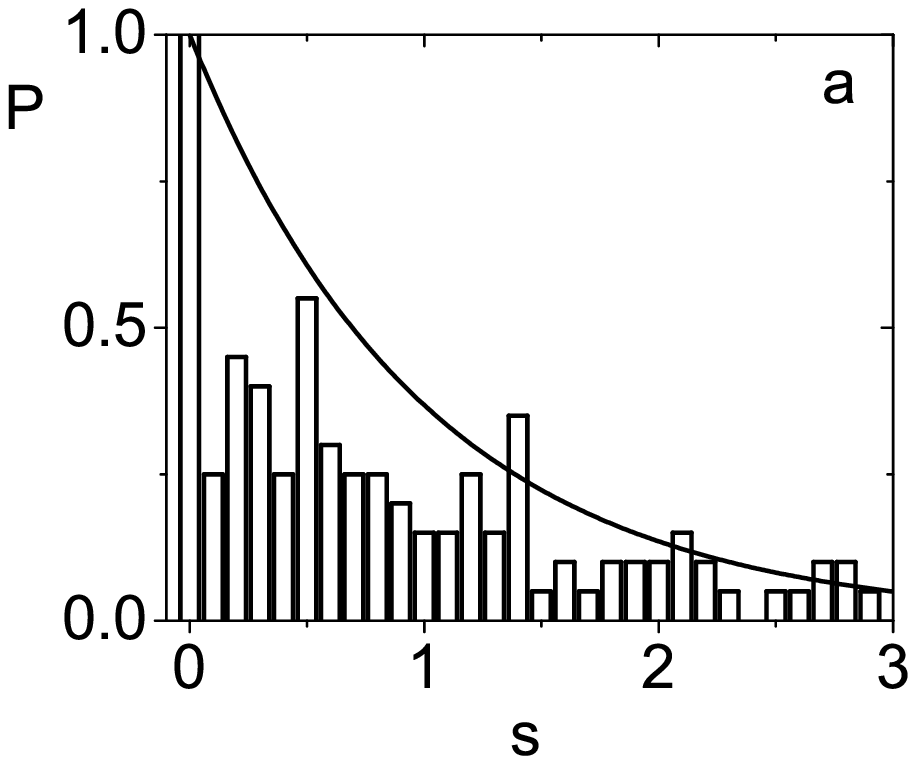}
\includegraphics[width=3.2in]{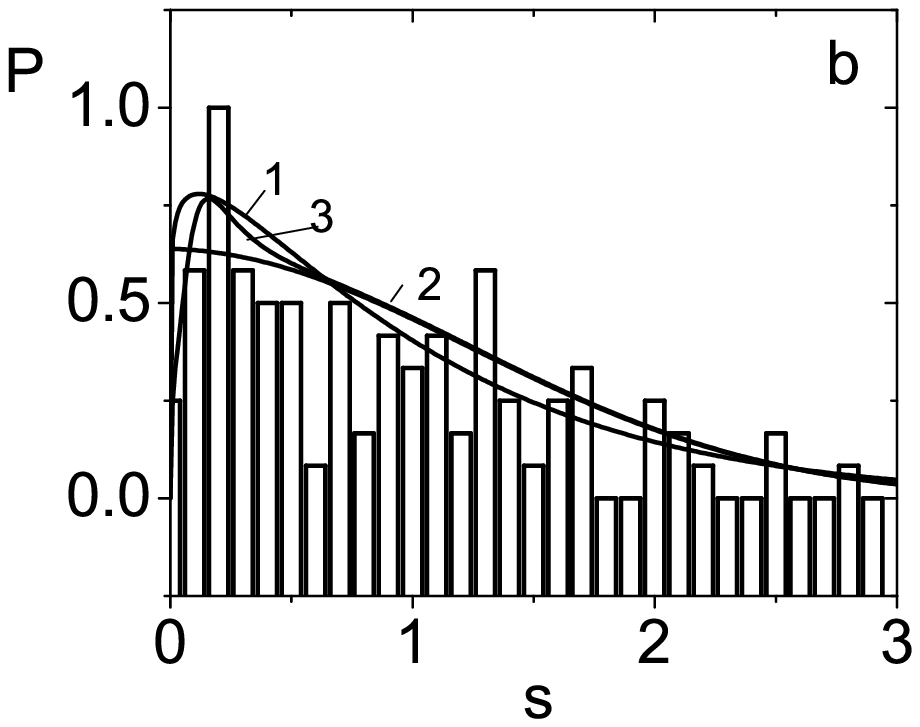}
\caption{\label{fig:epsart}The distribution of IF intervals, $P(s)$.
Figure 9a is for the empty resonator spectrum. Solid line is for the
Berry-Robnik distribution at $\rho  \to 1$ or the Brody distribution
at $\beta  \to 0$. Figure 9b is for the spectrum of the resonator
fully filled with styrofoam particles. The curve 1 is for Brody
distribution at $\beta  = 0.1$; the curve 2 is for Berry-Robnik
distribution at $\rho  = 0.4$; the curve 3 is for
Podolskiy-Narimanov distribution at $\rho  = 0.4$ and $\nu  = 0.1$
[27].}
\end{figure}

As we can see in Fig.9 the function $ P(s)$ has maximum at small $s$
that is not described by the Berry-Robnik dependence [26]. The
presence of such a maximum can be explained by Chaos-Assisted
Tunneling (CAT) [27] in the resonator with random inhomogeneities.
The proposed in [27] IF distribution function gives the possibility
to describe the observed distribution at the parameter $\rho  = 0.4$
corresponding to the classical dynamics in the system and the
parameter $\nu  = 0.1$ describing the tunneling between different
states of modes.

We calculated the spectral rigidity for the resonator with
inhomogeneities. The spectral rigidity $\Delta _3 (L)$  is an
integral characteristic of degree of spectral lines ordering for
frequency distances that is much more than inter-frequency interval.
It is given by [15]

\begin{equation}
\Delta _3 (x,L) = \frac{1}{L}\min _{A,B} \int\limits_x^{x + L}
{[n(\varepsilon ) - A\varepsilon  - B]^2 d\varepsilon }
\end{equation}

where $L$ is the interval length on which the function $\Delta _3
(x,L)$  is determined. The function ${n(\varepsilon )}$ is built as
follows [15]. For the sequence of the frequencies $\omega _n^{} $
normalized on unit density ($\omega _n^{}  = \omega _{n - 1}^{}  +
S_n $), we introduce a staircase function $n(\varepsilon )$ equals
the number of frequencies with $\omega _n^{}  < \varepsilon $.

The function $n(\varepsilon )$ has a staircase view with average
unit tilt. The function $\Delta _3 (x,L)$ is determined as the
minimum of quadratic deviation $n(\varepsilon )$ in the interval
$(x,\,\;x + L)$ from the straight line. The meaning of the averaged
in $x$ spectral rigidity (Eq.(34)) $\left\langle {\Delta _3 (x,L)}
\right\rangle _x $ depends only on $L$ and is denoted as $\Delta _3
(L)$.

The obtained curve $\Delta _3 (L$ for the resonator with random
inhomogeneities is shown in Fig.10. This curve (2) is placed between
the spectral rigidity for the Poisson distribution (curve 1,
$(\Delta _3 (L) = L/15)$)) and curve 3 that is corresponded to the
spectral rigidity for Gauss orthogonal ensemble (GOA) $\Delta _3 (L)
= 1/\pi ^2 \ln L - 0.00687$ [15]. The latter is observed at the
modeling of quantum chaos in microwave cavity unstable resonators
similar to Sinai and Bunimovich billiards [14].

\begin{figure}
\includegraphics[width=3.2in]{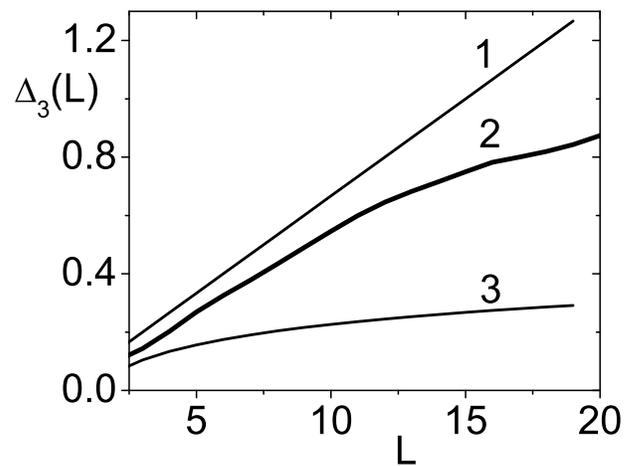}
\caption{\label{fig:epsart} The spectral rigidity  for the resonator
filled with styrofoam particles (curve 2). The straight line is for
the spectral rigidity of Poisson distribution (curve 1), curve 3 is
for GOA.}
\end{figure}

\section{Active resonator with random inhomogeneities}
We studied also an active quasi-optical resonator with random
inhomogeneities.  In comparison with the considered above passive
resonator the active one is a self-sustained oscillation system
where the presence of random inhomogeneities affects on the
excitation of resonator oscillations. For that we used the same
resonator as mentioned above (Fig. 4) with pointed Gunn diode
inside. The microwave electrical field for $TE$ mode was directed
along diode's axis. The diode DC power supply was implemented
through a filter as a quarter-wave microwave isolator; owing to
spurious microwave radiation was prevented.

Quasi-optical resonator with Gunn diode is an active oscillator with
distributed parameters. Near threshold of excitation in such an
oscillator with the empty resonator was detected unstable
multi-frequency generation. Such kind of generation we can explain
by frequency jumps between adjacent spectral lines with high quality
factor. If excitation threshold was highly exceeded, as a result of
frequency competition, corresponding to these spectral lines,
mono-frequency generation occurs (Fig.11b). The active oscillator
selects "itself" the only frequency to provide maximal regeneration
factor [28].

The random inhomogeneities lead to effective rarefaction of the
resonator spectrum. The number of adjacent spectral lines with high
quality factor is reduced, and, as a result, multi-frequency
generation disappears. At small exceeding of the generation
threshold the noise generation appears (Fig.11c) in the resonator
with inhomogeneities. At big exceeding of the generation threshold
the stable mono-frequency is observed (Fig.11d). Owing to
inhomogeneities mono-frequency generation possesses greater
frequency stability, and the number of generating frequencies is
reduced in the range of diode negative resistance. Fig.12 shows
generating frequencies for the empty resonator and with random bulk
inhomogeneities one.

\begin{figure}
\includegraphics[width=1.5in]{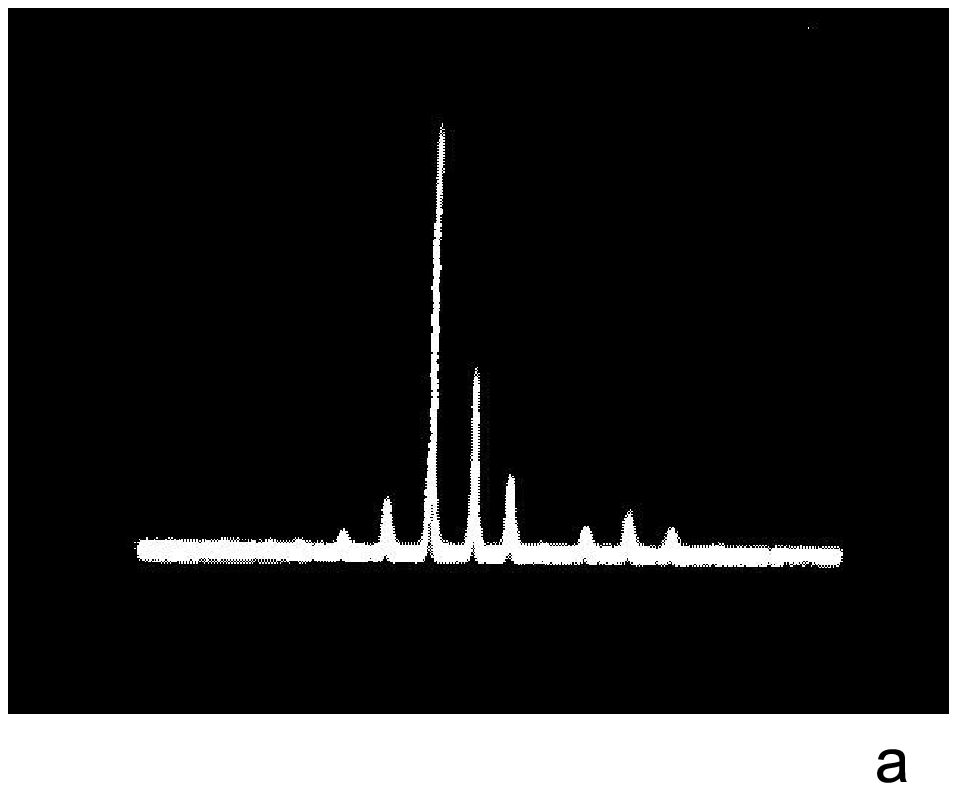}
\includegraphics[width=1.5in]{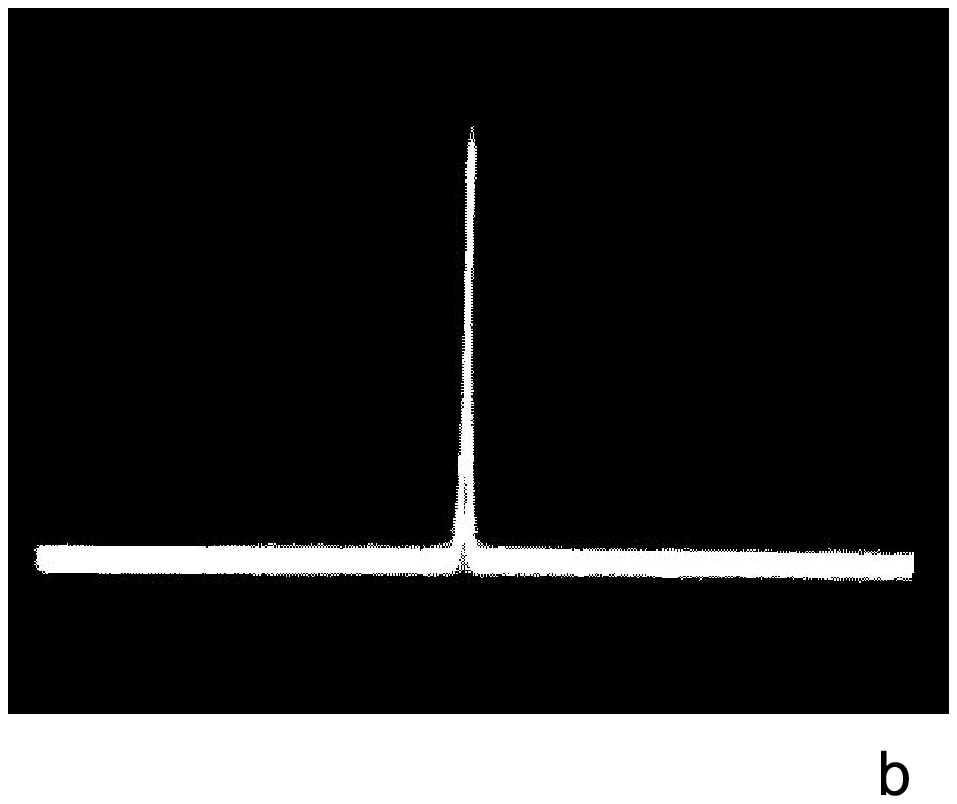}
\includegraphics[width=1.5in]{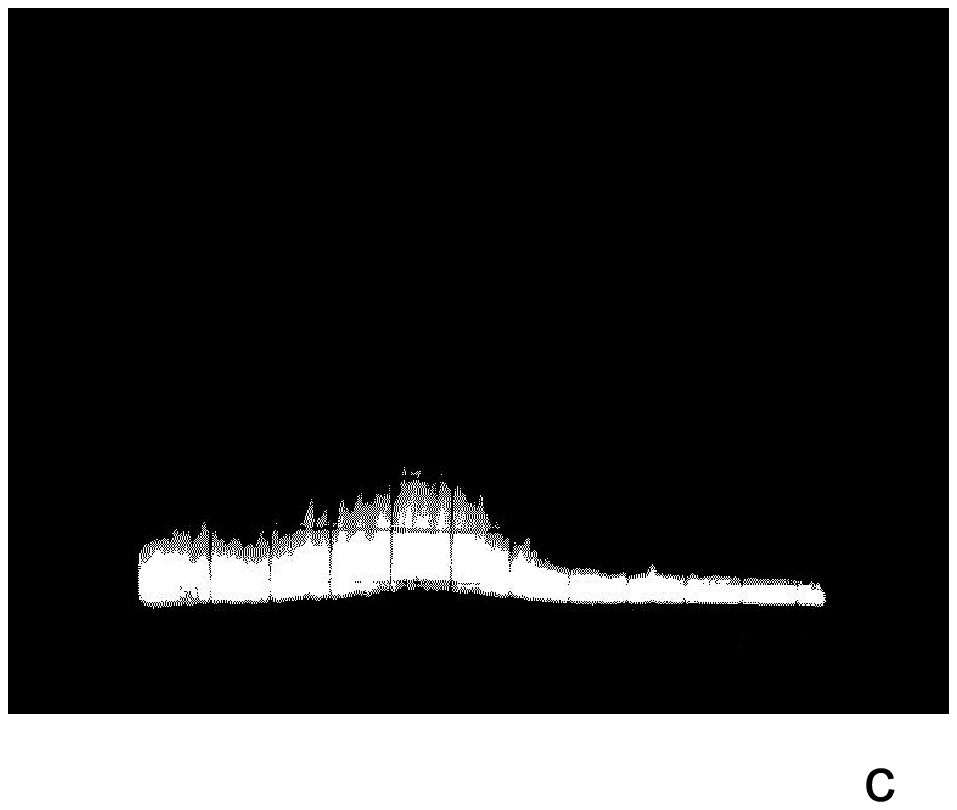}
\includegraphics[width=1.5in]{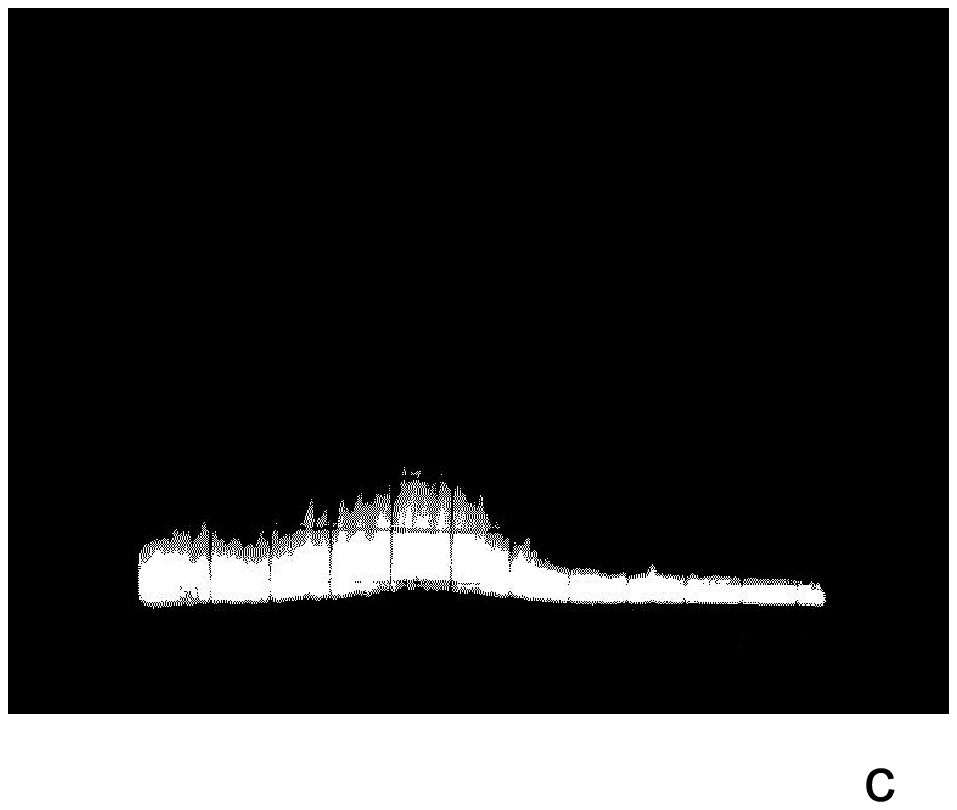}
\caption{\label{fig:epsart} Oscillogram of the 36 GHz generation
obtained by a millimeter wave spectrum analyzer. (a) and (b) are for
the empty resonator;  multi-frequency generation near the generation
threshold (a) and mono-frequency generation much far from the
generation threshold (b). (c) and (d) are for the resonator filled
with inhomogeneities; chaotic generation near the generation
threshold (c) and mono-frequency stable generation much far from the
generation threshold (d).}
\end{figure}

\begin{figure}
\includegraphics[width=3.2in]{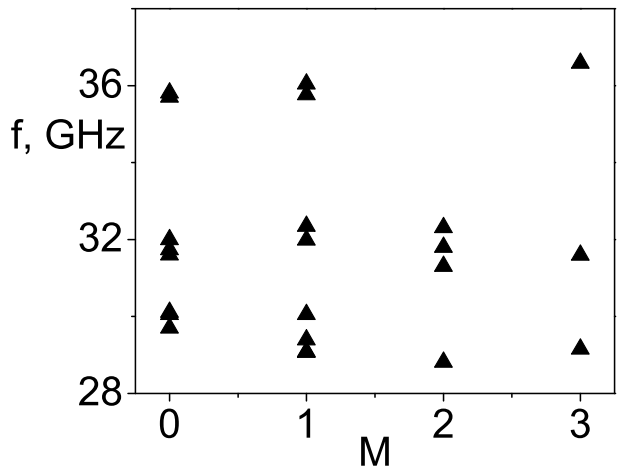}
\caption{\label{fig:epsart} The generating frequencies $f$ depending
on the number of inhomogeneities $M$. The empty resonator ($M = 0$),
the resonator quarter-filled with inhomogeneities ($ M = 1$), the
resonator half-filled with inhomogeneities ($ M = 2$), the resonator
fully filled with inhomogeneities ($M = 3$). }
\end{figure}

\section{Conclusion}
The statistical spectral theory of quasi-optical cavity resonator
filled with random dielectric inhomogeneities was developed in the
present paper. We showed that the presence of inhomogeneities leads
to the broadening and shifting of spectral lines. It is found out
that the nature of broadening and shifting of spectral lines is
relevant to the intermode scattering. The scattering effect for the
given spectral line essentially depends on frequency distance
between of it and adjacent ones and is sharply decreased for bigger
distances. Under the influence of ransom inhomogeneities original
spectrum modification occurs that can be interpreted as spectrum
rarefaction. The spectrum is rarefied because of solitary spectral
lines are not practically subjected to the influence of
inhomogeneities. The quality factor of such lines and,
correspondingly, their intensity stay high at the resonator
excitation. The intensity of adjacent lines broadened under the
influence of inhomogeneities is essentially reduced. Owing to that
at the great number of inhomogeneities the resonator spectrum is
"rarefied" - few solitary high quality factor spectral lines prevail
in the spectrum. Theoretical prediction of broadening and shifting
of resonator spectral lines and spectrum  "rarefaction" are subject
to experimental check.

For that purpose we studied experimentally  in 8-millimeter wave
range the spectrum of the quasi-optical cavity resonator filled with
random small-scattered bulk inhomogeneities. These inhomogeneities
were styrofoam particles with smaller size than operating
wavelength. It is found out that such inhomogeneities lead to
broadening and shifting the spectral lines. As experiment showed the
maximum of their influence was on frequency adjacent spectral lines.
The solitary lines, according to our theory, were subjected to this
influence in much smaller degree. We detected the effect of
stochastic spectrum "rarefaction". They prove that the main
mechanism of broadening and shifting of spectral lines is relevant
to inhomogeneities intermode scattering. We studied also chaotic
properties of oscillations in our resonator. It is found out that
the empty resonator has IF intervals distribution similar to Poisson
distribution that is typical to the spectrum with non-correlated IF
intervals. Even at small number of inhomogeneities the resonator
spectrum has random part that increases in proportion to the number
of inhomogeneities in the resonator. At that IF intervals
distribution in random inhomogeneous resonator is described by Brody
and Berry-Robnik distributions of IF intervals.

We obtained the results concerning the influence of bulk
inhomogeneities on the process of generation in a self-oscillatory
system.  The self-oscillatory system was a quasi-optical millimeter
wave cavity resonator containing inhomogeneities with an active
element as Gunn diode. We detected that inhomogeneities leads to the
essential "rarefaction" of the spectrum and creates conditions for
monochromatic stable generation in self-oscillatory system. The
inhomogeneous quasi-optical cavity millimeter wave resonator
(passive and active) can serve as a model of semiconductor quantum
billiard. Based on our results we suggest using such billiards with
spectrum rarefied by random inhomogeneities as an active system of
semiconductor laser.


\begin{thebibliography}{99}
\bibitem{1}
S.M. Rytov, Ya.A. Kravtsov, and V.I. Tatarskii, \emph{Principles of
Statistical Radiophysics} (Springer Verlag, V.4. Berlin,  1987).
\bibitem{2}
F.G.Bass, I.M.Fuks,  \emph{Wave Scattering from Statistically Rough
Surfaces} (Pergamon Press, N.Y.,  1979).
\bibitem{3}
E.M. Ganapolski, Z.E. Eremenko, Phys. Rev. E, \textbf{65}, 056218
(2002).
\bibitem{4}
V.B. Braginskii, V.P. Mitrofanov, and V.I. Panov,  \emph{Systems
with low dissipations}  (Nauka, Moscow,  1981).
\bibitem{5}
A.N.Oraevskii, M.Scally, and V.L. Velichanskii,  Quantum
Electronics, \textbf{25}, 211 (1998).
\bibitem{6}
B.E.Little, J-P. Laine, S.Chu, Optics Letters, \textbf{22}, 4
(1997).
\bibitem{7}
L. Goldstein, F. Glass, J.Y. Marzin,  M.N. Charasse,  and G.Le.
Roux, Appl. Phys. Lett., \textbf{47}, 1099 (1985).
\bibitem{8}
P.M.Petroff, S.P. Den Baars, Superlat. Microstr., \textbf{15}, 15
(1994).
\bibitem{9}
M. Moison, F. Houzay, F. Barthe, L. Leprince, E.Andre, O. Vatel,
Appl. Phys. Lett., \textbf{64},  196 (1994).
\bibitem{10}
Zh.I. Alferov et al, Fizika i Tekhnika Poluprovodnikov [Soviet
Physics -- Semiconductors], \textbf{30}, 351 (1996).
\bibitem{11}
Zh. I. Alferov et al, Fizika i Tekhnika Poluprovodnikov [Soviet
Physics -- Semiconductors], \textbf{30}, 357 (1996).
\bibitem{12}
Yu. M. Schernjakov et al, Pis'ma v Zhurnal Tekhnicheskoi Fiziki
[Soviet Technical Physics Letters], \textbf{23}, 51 (1997).
\bibitem{13}
V.I. Beljavskii, S.V.Shevtsov, Fizika i Tekhnika Poluprovodnikov
[Soviet Physics -- Semiconductors], \textbf{36}, 874 (2002).
\bibitem{14}
H.J. Stockmann, \emph{Quantum Chaos An Introduction} (Cambridge
University Press, 1999).
\bibitem{15}
P.V. Elutin, Uspekhi Fizicheskikh Nauk [Soviet Physics -- Uspekhi],
\textbf{155}, 397 (1988).
\bibitem{16}
H. Alt, H. D.Graf, R. Hofferbert et al. Phys. Rev. E, \textbf{54},
2303 (1996).
\bibitem{17}
E.M.Ganapolskii, Z.E.Eremenko, Dopovidi Akademii Nauk Ukrayiny
[Ukraniain Physics -- Doklady], No.12, 93 (2000).
\bibitem{18}
Yu.V. Tarasov, Waves in Random Media, \textbf{10}, 395 (2000).
\bibitem{19}
Yu.V. Tarasov, Low Temp. Phys., \textbf{29}, 45 (2003).
\bibitem{20}
L.A. Vainstein,  \emph{Electromagnetic waves} (Radio i  Svjaz,
Moscow, 1988).
\bibitem{21}
Yu.V. Tarasov, Phys. Rev. B, \textbf{71}, 125112 (2005).
\bibitem{22}
Yu.V. Tarasov, Phys. Rev. B, \textbf{73}, 014202 (2006).
\bibitem{23}
R. Newton. {\it Scattering Theory of Waves and Particles}
(McGraw-Hill, New-York, 1968).
\bibitem{24}
I.M. Lifshitz, S.A. Gredeskul, L.A. Pastur,  \emph{Intriduction in
Theory of Disordered Systems} (Nauka, Moscow, 1982).
\bibitem{25}
I.S. Gradshtein, I.M. Ryzhik, \emph{Tables of Integrals, Sums,
Series and Products} (Nauka, Moscow, 1971).
\bibitem{26}
M.V. Berry, M. Robnik, J. Phys., \textbf{A17}, 2413 (1984).
\bibitem{27}
V.A. Podolskiy, E.E. Narimanov, Phys. Rev. Lett., \textbf{91},
263601 (2003);  V.Podolskiy, E.Narimanov, arXiv: nlin.CD/0310034.
\bibitem{28}
D.P. Tzarapkin,  \emph{Microwave Gann oscillators} (Radio i Svjaz,
Moscow,  1982).
\end{thebibliography}

\end{document}